\documentclass[12pt]{article}
 \usepackage[dvips,final]{graphicx}
  \usepackage{bm}
   \usepackage{amsmath}
    \usepackage{amssymb}
\newcommand{\be}{\begin{equation}}
\newcommand{\ee}{\end{equation}}
\newcommand{\ba}{\begin{eqnarray}}
\newcommand{\ea}{\end{eqnarray}}
\newcommand{\baa}{\begin{eqnarray*}}
\newcommand{\eaa}{\end{eqnarray*}}
\newcommand{\eb}{\end{thebibliography}}

\newcommand{\ci}[1]{\cite{#1}}
\newcommand{\bi}[1]{\bibitem{#1}}
\begin{document}
\begin{flushright}
CPT-2004/P.017 \\

\end{flushright}

\bigskip\bigskip

\begin{center}
{\Large \bf QCD radiative and power corrections and 
Generalized GDH Sum Rules}\\[1cm]
{ \bf J. Soffer $^{a}$ and O.Teryaev$^{a,b}$}\\[0.3cm]
$^a$Centre de Physique Th\'eorique \footnote {{UMR 6207 - Unit\'e Mixte de Recherche 
du CNRS et des Universit\'es Aix-Marseille I, Aix-Marseille II et de
l'Universit\'e du Sud Toulon-Var - Laboratoire affili\'e à la FRUMAM.}}, CNRS-Luminy,\\
Case 907, F-13288 Marseille Cedex 9, France \\ 
\vskip 0.5cm
$^b$Bogoliubov Laboratory of Theoretical Physics,\\
Joint Institute for Nuclear Research\\
141980 Dubna, Moscow region, Russia\footnote{Permanent address}
\end{center}

\begin{abstract}

We extend the earlier suggested QCD-motivated
 model for the $Q^2$-dependence of the generalized Gerasimov-Drell-Hearn (GDH)
 sum rule which assumes the smooth dependence
 of the structure function $g_T$, while the sharp dependence is due
 to the $g_2$ contribution and is described by the elastic part
 of the Burkhardt-Cottingham sum rule. The model successfully predicts the
 low crossing point for the proton GDH integral, but is at variance with the
 recent very accurate JLAB data. We show that, at this level of accuracy, one
 should include the previously neglected
 radiative and power QCD corrections, as boundary values for the model.
 We stress that the GDH integral, when measured with such a high accuracy
 achieved by the recent JLAB data, is very sensitive to QCD power corrections.
 We estimate the value of these power corrections from the JLAB data
 at $Q^2 \sim 1 \mbox{GeV}^2$.
 The inclusion of all QCD corrections leads to a good description of proton, 
 neutron and deuteron 
 data at all $Q^2$.

\end{abstract}

\bigskip


\newpage

The generalized ($Q^2$-dependent) Gerasimov-Drell-Hearn
(GDH) sum rules
\cite{Ger,DH} 
are just being tested experimentally for proton,
neutron and deuteron \cite{E143,JLAB,Hermes,JLAB2}.
The characteristic feature of the proton data is the strong dependence 
on the four-momentum transfer $Q^2$,
for $Q^2 <1 \mbox{GeV}^2$, with a zero crossing for $Q^2 \sim 200-250 \mbox{MeV}^2$,
which is in complete agreement
with our prediction \cite{ST93,ST95}, published almost 10 years ago.
Our approach is making use of the relation to the Burkhardt-Cottingham
sum rule for the structure function $g_2$, whose elastic contribution
is the main source of a strong $Q^2$-dependence, while
the contribution of the other structure function, $g_T=g_1+g_2$ is smooth.

However, the recently published proton JLAB data \cite{JLAB2} lie below
the prediction, displaying quite a similar shape.
Such a behaviour suggests, that
the reason for the discrepancy may be the oversimplified 
treatment of the QCD expressions at the boundary point $Q_0 \sim 1 \mbox{GeV}$,
defined in the smooth interpolation between large $Q^2$ and $Q^2=0$ and
which serve as an input for our model. For large $Q^2$ we 
took the asymptotic value for the GDH integral and we 
neglected all the calculable corrections, as well as the 
contribution of the $g_2$ structure function. 
This was quite natural and unnecessary 10 years ago, since no data 
was available at that time.  

In the present paper we fill this gap and include the radiative (logarithmic) and power
QCD corrections. 
We found that the JLAB data are quite sensitive to power corrections
and may be used for the extraction of the relevant phenomenological 
parameters. 
We present here the numerical values of these parameters which naturally 
depend on the approximation of the QCD  
perturbation theory. The resulting theoretical uncertainty should be of order of the last term  
of the perturbative series , taken into account, and should not therefore be more than several percents.

Moreover, the perturbative series should contain the renormalon ambiguity due to the factorial growth of the coefficients, 
resulting in a power rather than logarithmic correction with an unspecified coefficient.   
It is in fact this ambiguity which allows the interpretation of the dependence of the numerical value of the power correction, 
earlier mentioned for 
the case of the $F_3$ structure function \cite{KatSid}, as an ambiguity in the 
separation of logarithmic and power corrections.  

We use the values of the power corrections as an input for our model at 
$Q_0^2 \sim 1 \mbox{GeV}^2$ 
and we achieve a rather good description of the proton data at lower $Q^2$.
We also present the improved description of the
neutron and deuteron data and the behaviour of the Bjorken sum rule at low $Q^2$.

The starting point of our approach is the analysis of the 
general tensor structure of $W^{\mu \nu}_A$, the spin-dependent part of 
hadronic tensor $W^{\mu \nu}$.
It is a linear combination of all possible Lorentz-covariant  
tensors, which should be orthogonal to the virtual photon momentum $q$, as
required by gauge invariance, and linear in the nucleon
covariant polarization $s$, from a general property
of the density matrix. If the nucleon has momentum $p$, we have as
usual,
$s\cdot p=0$ and $s^2=-1$. There are only two such tensors: the first one
arises already in the Born diagram
\be T_1^{\mu
\nu}= \epsilon^{\mu \nu \alpha \beta}
s_\alpha q_\beta
\end{equation}  
and the second
tensor is just
\begin{equation} T_2^{\mu \nu}=(s\cdot q) \epsilon^{\mu \nu \alpha\beta}p_{\alpha}q_{\beta}~.
\end{equation}  
The scalar coefficients of these tensors are specified in a
well-known way, since we have
\ba
W^{\mu \nu}_A={-i\epsilon^
{\mu \nu  \alpha\beta}\over {p\cdot q}}q_{\beta}
(g_1 (x, Q^2)s_{\alpha}+g_2 (x,Q^2)(s_{\alpha}-p_{\alpha}{s\cdot q \over
{p\cdot q}}))=\nonumber \\
{-i\epsilon^ {\mu \nu  \alpha\beta}\over {p\cdot q}}q_{\beta}((g_1 (x,
Q^2)+g_2 (x,Q^2))s_{\alpha}-g_2
(x,Q^2)p_{\alpha}{s\cdot q \over {p\cdot q}})~. 
\ea
Therefore, due to the factor 
$(s\cdot q)$, $g_2$ is making the difference
between longitudinal and transverse polarizations, while
$g_T=g_1+g_2$ contributes equally in both cases.

Let us consider now the $Q^2$-dependent integral
\begin{equation} I_1(Q^2)={2 M^2\over {Q^2}} \Gamma_1(Q^2) \equiv {2 M^2\over {Q^2}} \int^1_0 g_1(x,Q^2) dx~.
\label{I1}
\end{equation}  
It is defined for {\it all} $Q^2$, and $g_1(x,Q^2)$ is the obvious
generalization for all $Q^2$ of the standard scale-invariant structure function $g_1(x)$.
Note that the elastic contribution at $x=1$ is not
included in the above sum rule. Then,
by changing the integration variable $x \to Q^2/2 M \nu$,
one recovers at $Q^2=0$ the integral over all energies of spin-dependent 
photon-nucleon cross-section, whose value is defined by 
the GDH sum rule \cite{Ger,DH}
\begin{equation}\label{5} I_1(0)=-{\mu_A^2 \over 4}~,
\end{equation}  
where $\mu_A$ is the nucleon anomalous magnetic moment in nuclear magnetons.
While $I_1(0)$ is always negative, its value at large $Q^2$ is determined
by the $Q^2$ independent integral $\int^1_0 g_1(x) dx$, which is
positive for the proton and negative for the neutron.

The separation of the contributions of $g_T$ and $g_2$
leads to the decomposition of  $I_1(Q^2)$ as the
difference between $I_{T}(Q^2)$ and $I_2(Q^2)$
\begin{equation} I_1(Q^2)=I_{T}(Q^2)-I_2(Q^2),
\end{equation}  
where
\begin{equation} I_{T}(Q^2)={2 M^2\over {Q^2}} \int^1_0 g_{T}(x,Q^2) dx,
\;\;\;\;I_2(Q^2)={2 M^2\over {Q^2}} \int^1_0 g_2(x,Q^2) dx~.
\end{equation}  

There are solid theoretical arguments to expect a strong $Q^2$-dependence
of $I_2(Q^2)$. It is the well-known Burkhardt-Cottingham sum rule \ci{BC}.
It states that
\begin{equation} \label{el}
I_2(Q^2)={1\over 4}\mu G_M (Q^2)
\frac{\mu G_M (Q^2) - G_E (Q^2)}{1+Q^2/4M^2},
\end{equation}  
where $\mu$ is the nucleon magnetic moment, $G_M (Q^2)$ and $G_E (Q^2)$ denote the familiar
Sachs form factors, which are dimensionless and normalized to unity
at $Q^2=0$, $G_M (0) = G_E (0) =1$. For large $Q^2$, as a consequence of 
the $Q^2$ behavior of the r.h.s.
of Eq.~(\ref{el}), we get
\begin{equation}\label{BC}
\int^1_0 g_2(x,Q^2) dx=0.
\end{equation}  

In particular, from Eq.~(\ref{BC}) it follows that
\begin{equation} I_2(0)={\mu_A^2+\mu_A e \over 4},
\end{equation}  
$e$ being
the nucleon charge in elementary units.
To reproduce the GDH value (see Eq.~(\ref{5})) one should have
\begin{equation} I_{T}(0)={\mu_A e \over 4},
\end{equation}  
which was indeed proved by Schwinger \cite{Sch}.
The importance of the $g_2$ contribution can be seen
already, since the entire $\mu_A$-term for the GDH sum rule is
provided by $I_2$.

Note that $I_{T}$ does not differ from $I_1$ for large $Q^2$ due to the
BC sum rule, but it is {\it positive} in the proton case.
It is possible
to obtain a smooth interpolation for  $I^p_{T}(Q^2)$
between large $Q^2$ and $Q^2=0$ \ci{ST93}

\begin{equation} I^p_{T}(Q^2)=\theta(Q^2_0-Q^2)({\mu_{A,p} \over 4}- {2 M^2 Q^2\over
{(Q^2_0)^2}} \Gamma^p_1)+\theta(Q^2-Q^2_0) {2 M^2\over {Q^2}}
\Gamma^p_1,
\end{equation}  
where $\Gamma^p_1=\int^1_0 g^p_1(x) dx$.
The continuity of the function and of its derivative is guaranteed
with the choice $Q^2_0=(16M^2/\mu_{A,p}) \Gamma^p_1 \sim \mbox{1GeV}^2$,
where the integral is given by the world average proton data.

This smooth interpolation seems to be very reasonable in the
framework of the QCD sum rules method \cite{ST95,ST02},
as the low energy theorem for the quantity linear in $\mu_A$
may, in principle, be obtained by making use of Ward identities.
It is also
compatible with resonance approaches \cite{Ioffe},
 as we observed
earlier \cite{ST95} that the magnetic transition to
$\Delta(1232)$, being the main origin
of sharp dependence in that approach, contributes only to
$g_2$.

However, such interpolation neglects 
the QCD perturbative and power corrections and on the other hand, it  
 assumes that at the boundary point 
$Q_0$ the contribution of $g_2$ is already extremely small
so that 
\begin{equation} I^p_{T}(Q_0^2)=I^p_{1}(Q_0^2).
\end{equation}  
 Both types of corrections are easily taken into account,
although this does not allow a simple analytic parametrization.

The starting point of the upgraded  model is the corrected 
expression for the asymptotic expression for $I_1^i$ ($i=p,n$):

\begin{equation} I_1^i(Q^2)={2 M^2\over {Q^2}} [\int^1_0 g_1^i(x,Q^2) dx (1- 
\frac{\alpha_s(Q^2)}{2 \pi})-c_i \frac{<<O_i>>}{Q^2}]~,
\label{I1c}
\end{equation}  
where we took into account the one-loop perturbative correction 
(while the inclusion of higher order ones will be discussed later), 
as well as the twist-4 contribution \cite{Power}. Here $c_i$ is the charge factor 
equal to $2/9$ for proton and to $1/18$
for neutron, while the matrix elements of the combinations of reduced twist-3 and -4 
operators happen to be equal \cite{Power} for both proton and neutron:
$$<<O_p>>=<<O_n>>=0.09 \pm 0.06 \mbox{GeV}^2~.$$
 
Note that the kinematical target mass corrections happen to be numerically small and 
we neglect their contribution. Anyway, they may be combined with the genuine twist corrections
and the resulting change of the latter is within experimental and theoretical errors.

As the expression for the $I_2$ stays unchanged, the expression for 
$I_T$ above the matching point $Q_0$ should change accordingly.
Let us start from the proton case.
\begin{equation} \label{asc}
I^p_{T,pert}(Q^2)=\theta (Q^2-Q^2_0){2 M^2\over {Q^2}} [\Gamma^p_1(1- 
\frac{\alpha_s(Q^2)}{2 \pi})-c_p \frac{<<O_p>>}{Q^2}-I^p_2(Q^2)].
\end{equation}  
The smooth interpolation to the GDH value at $Q^2=0$ is now more difficult and cannot be performed 
anymore, by making use of simple analytic formulae. Instead, we expand (\ref{asc}) to the
power series at the point $Q_0$ and define the expression at the low $Q^2$ as: 
\begin{equation} \label{npc}
I^p_{T,non-pert}(Q^2)= \theta (Q_0^2-Q^2)
\sum_{n=0}^N \biggl(\frac{1}{n !} \frac{\partial^n I^p_{T,pert}}{\partial 
(Q^2)^n}\biggr)_{Q=Q_0} (Q^2-Q^2_0)^n.    
\end{equation}  
Here $N$ is the number of continuous derivatives of these two expansions, which
turns out to be    
a free parameter of the model, together with the matching value $Q_0$. 
They should be chosen in such a way,
that the condition for real photons 
\begin{equation} \label{np0}
I^p_{T,non-pert}(0)= \frac{\mu_A}{4}
\end{equation}    
is satisfied. 

The procedure we are implementing in such a way, may 
be considered as a matching of 
the ``twist-like''
expansion in negative powers
of $Q^2$ and the ``chiral-like'' expansion in positive powers of $Q^2$, 
which is similar
to the matching of the expansions in
direct and inverse coupling constants. In its simplest present version
we take only the value $I(0)$ as an input, although the slope and other derivatives calculated within 
the chiral perturbation theory may be added in future work.

As soon as the low $Q^2$ region exhibits the important contribution of the resonances  \cite{Ioffe},
the suggested procedure may be also considered as a version of quark-hadron duality. 
It is worthy to note here, that Bloom-Gilman duality for spin-dependent case is 
strongly violated by the contribution of $\Delta(1232)$ resonance \cite{simula}. As it was 
mentioned before, since this resonance does not contribute to the $g_T$ structure function \cite{ST95}, it is this function
which may be a good candidate to study duality.

We have studied Eq.~(\ref{np0}) numerically changing the following inputs:

i) for different order of perturbative correction (1,2,3 loops) \cite{Pert}.

ii) for different values of the degree of approximating polynomial $N$ in Eq.~(\ref{npc});
it is interesting that taking $N=1$ does not allow for solution of Eq.~(\ref{np0}).  

iii) for different values of non-perturbative corrections, which 
we were choosing in order to 
be close to JLAB data at their highest $Q^2 \sim 1 \mbox{GeV}^2$. 
We observed that the increasing of the order 
of perturbative corrections lead to systematical decrease of the required non-perturbative one, 
which is similar to the case of $F_3$ structure function \cite{KatSid} and 
may be considered as a manifestation 
of the ambiguity in separating logarithmic and power corrections. 

iv) we varied the matching point $Q_0$ until Eq.~(\ref{np0}) is satisfied. 

We found that $Q_0$ is systematically (but not strongly) increasing with $N$. 
The expression for $Q^2$ dependent integral $\Gamma_1^p(Q^2)=I^p_{1,non-pert}(Q^2) Q^2/2M^2$ 
resulting from 
3-loops perturbative correction with $N=3, <<O_p>>=0.11 \mbox{GeV}^2$ and $Q_0^2 =0.97 \mbox{GeV}^2$ 
is shown in Fig.~1. 
It is reasonably close to the JLAB data \cite{JLAB2}.
In what follows the thick lines correspond to our new approach and we present also the results from the 
old approach for comparison. 

\begin{figure}[ht]
 \centerline{\includegraphics[width=0.88\textwidth]{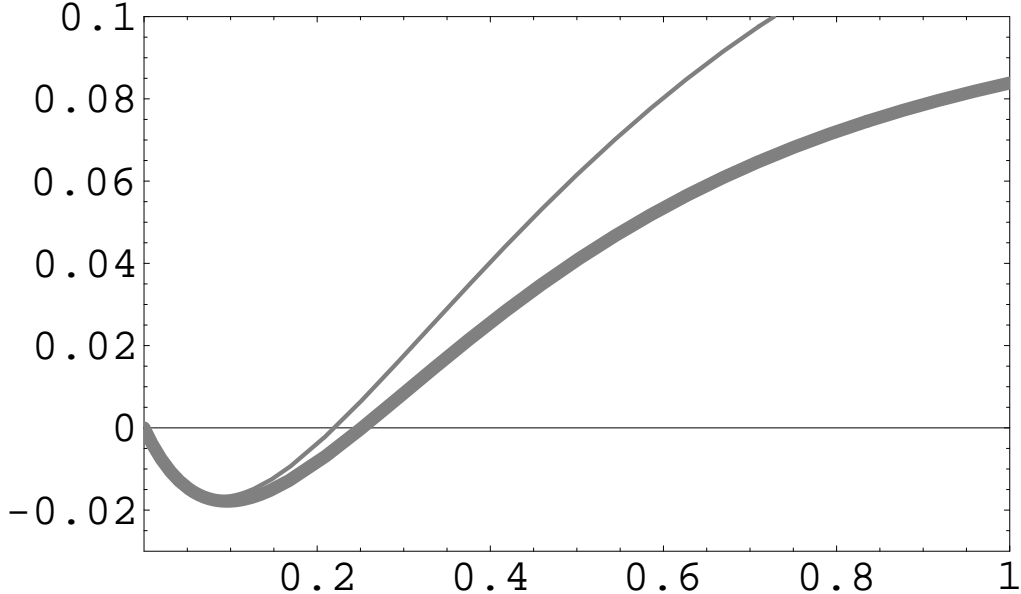}}
 \caption[*]{\baselineskip 1pt
  Our description of $\Gamma^{p}_1(Q^2)$ versus $Q^2$. The thick line is the new analysis to be
  compared with the thin line, which represents our previous approach without corrections.
  \label{Fig1}}
\end{figure}

Here we took the asymptotic value for $\Gamma_1^p=0.147$ providing the good description of
$\Gamma_1^p(Q^2)$ for $Q^2$ of the order of several $GeV^2$, when the 3-loops radiative correction is included.
This procedure may be considered as a sort of preliminary estimate, since the full 3-loops analysis is
not available.

To generalize our approach to the neutron case, we use the difference between proton and
neutron instead of the neutron itself. Although it is possible, in principle,
to construct a smooth interpolation for the functions $g_1$ themselves
\ci{book}, it does not fit the suggested general argument
 on the
linearity in $\mu_A$, since 
$I^{p-n}_{1}(0)$ is proportional to $\mu_{A,n}^2- \mu_{A,p}^2$, which
is quadratic and, moreover,
has an additional suppression due to the smallness
of isoscalar anomalous magnetic moment.
So we suggest the following parametrization for the isovector
contribution of $I_{T}(Q^2)$, namely $I^{p-n}_{T}(Q^2)$, 
above the matching point, where again only the 1-loop term is presented explicitly  
\begin{eqnarray}
\label{nc}
I^{p-n}_{T,pert}(Q^2)=\theta (Q^2-Q^2_1){2 M^2\over {Q^2}} [\Gamma^{p-n}_1(1- 
\frac{\alpha_s(Q^2)}{2 \pi}) \nonumber \\
- c_p \frac{<<O_p>>}{Q^2} + c_n\frac{<<O_n>>}{Q^2} -I^p_2(Q^2)+I_2^n(Q^2)].
\end{eqnarray}

Here the transition value
$Q_1^2$ may be determined by the continuity
conditions in a similar way. We get the
value $Q_1^2 \sim 1.04 \mbox{GeV}^2$, which is not too far  
from that of the proton case. Concerning $I_2^n(Q^2)$, which is given by Eq.~(\ref{el}),
we have not neglected $G_E^n(Q^2)$ \footnote{ We thank G.Dodge for making this suggestion.
} and we have used its very recent
 determination \cite{Mad}.
The result of this calculation is very slightly modified compared to the case where one
assumes
$G_E^n(Q^2)=0$ and all the subsequent results involving the neutron were obtained
with a non-zero $G_E^n(Q^2)$.
The asymptotic value of $\Gamma_1^{p-n}=0.21$ is dictated by the Bjorken sum rule.
We also took the same value $<<O_n>>=0.11 \mbox{GeV}^2$ 
as in the proton case.  
The plot representing $\Gamma^{p-n}_1 (Q^2)$ is displayed
on Fig.~2 and agrees well with the very recent experimental data \cite{Deur}. 
 
This may be considered as an argument in favour of
the general picture of power corrections obtained
in QCD sum rules calculations \cite{Power},
where the neutron correction is small. However the
quantitative comparison with
the calculations in the framework of the chiral soliton
model \cite{sol}
would require a more detailed analysis.

Now we have all the ingredients to turn to the behavior of neutron integral, which is
simply obtained from the difference $\Gamma_1^p(Q^2) - \Gamma_1^{p-n}(Q^2)$. It is shown
on Fig.~3 and we notice that the strong oscillation around 
$Q^2 =1\mbox{GeV}^2$, we had in the previous analysis, is no longer there.

\begin{figure}[ht]
 \centerline{\includegraphics[width=0.88\textwidth]{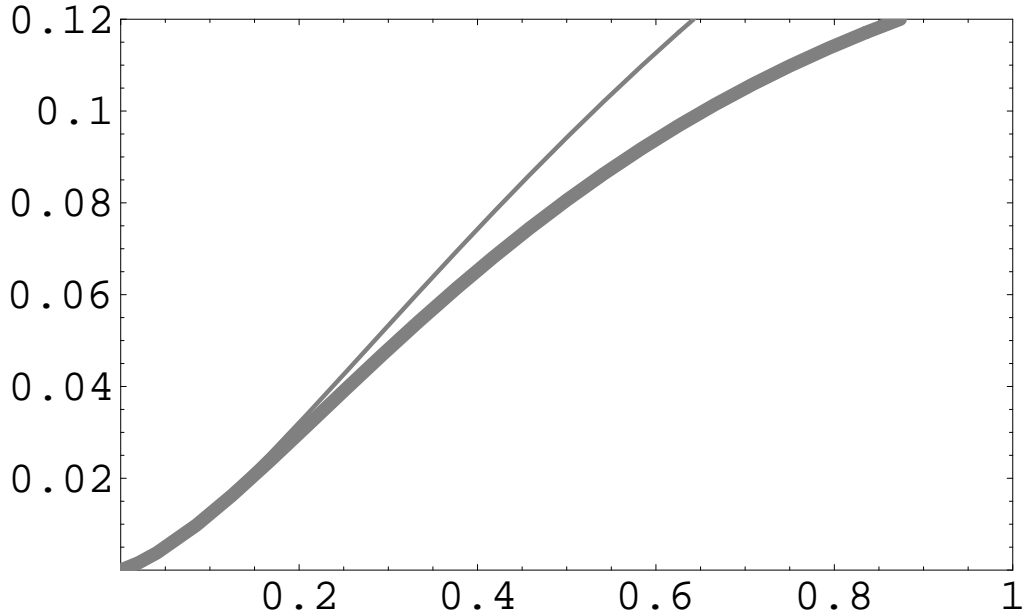}}
 \caption[*]{\baselineskip 1pt
Our prediction for $\Gamma_1^{p-n}(Q^2)$ versus $Q^2$. The thick line is the new analysis to be
compared with the thin line, which represents our previous approach without corrections.
 \label{Fig2}}
\end{figure}

\begin{figure}[ht]
 \centerline{\includegraphics[width=0.88\textwidth]{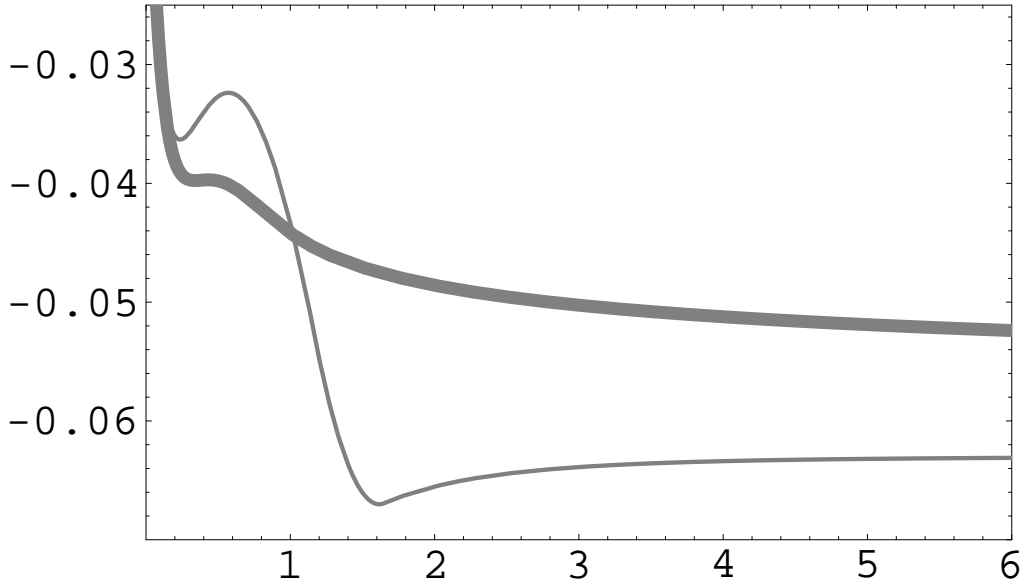}}
\caption[*]{\baselineskip 1pt
 Our prediction for $\Gamma_1^{n}(Q^2)$ versus $Q^2$. The thick line is the new analysis to be
compared with the thin line, which represents our previous approach without corrections.
 \label{Fig3}}
\end{figure}

\newpage

We now have all the ingredients to investigate the deuteron integral.
Note that in this case the generalization of the GDH sum rule may be naturally 
decomposed into two distinct regions. 

The first one is the region of large $Q^2$, where nuclear binding effects
can be
disregard, so that the deuteron structure function is 
the simple additive sum of the proton and neutron ones. As a result, the $Q^2 \to 0$ limit 
of this intermediate asymptotics is defined by the sum of the squares of proton 
and neutron anomalous magnetic momenta. 

\begin{figure}[ht]
 \centerline{\includegraphics[width=0.88\textwidth]{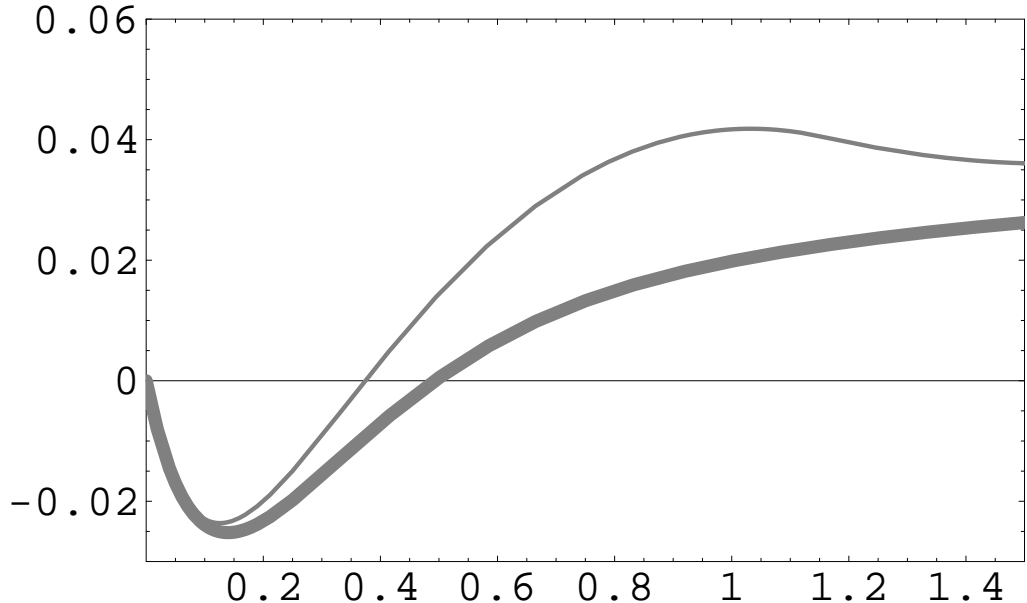}}
\caption[*]{\baselineskip 1pt
 Our prediction for $\Gamma_1^{d}(Q^2)$ versus $Q^2$. The thick line is the new analysis to be
  compared with the thin line, which represents our previous approach without corrections.
 \label{Fig4}}
\end{figure}

When nuclear binding effects are taken into account one should get instead the square of 
the sum of these anomalous magnetic moments.
As they are known to be rather close in magnitude and of different sign, the result
 should be therefore very small.  

The difference between these two regimes should be attributed \cite{ot03} to the deuteron 
photodesintegration channel, 
which is supported by existing explicit calculations \cite{aren} in the case of real
 photons. For virtual photons, 
this allows to estimate the $Q^2$ value, where binding effects start to play a role, to be 
of the order of $m_\pi^2$.
The simplest way to implement this reasoning \cite{ot03} is to use the following expression:

\begin{equation} 
I^d_{1}(Q^2)=\frac{1}{2}[\theta(Q_{d0}^2-Q^2)\frac{2 Q^2}{Q_{d0}^2+Q^2}+\theta(Q^2-Q_{d0}^2)]
(I^p_1(Q^2)+ I^n_1(Q^2)
\end{equation}  

Here we introduced the nuclear scale $Q_{d0}\sim m_\pi$ and neglected the square of the 
deuteron anomalous magnetic moment. The prediction is shown in Fig.~4 and 
  seems to be in good agreement with the preliminary JLAB data \cite{dodge}.

Let us finally discuss the role of the elastic contribution. It must be definitely included \cite{ji}
if one uses the operator product expansion (OPE),~ 
which is the essential tool in determining the power corrections. 
However, we use the OPE only above matching point, where the elastic contribution is small.
At the same time, below the matching point the object those $Q^2$ behaviour is studied
may be considered as a sort of fracture function, where only a partial summation over final states, 
excluding the elastic one, is implied. It is this function which may reach the GDH value at $Q^2=0$.\\\\

{\bf Acknowledgments}

We are indebted to Vladimir Braun, Volker Burkert, Claude Bourrely, Gail Dodge, Sebastian K\"uhn and Zein-Eddine 
Meziani for discussions 
and correspondence. 
O.T. is indebted to CPT for warm hospitality and financial support.
He was also partly supported by RFBR (Grant 03-02-16816) and by INTAS (International
Association for the Promotion of Cooperation with Scientists from the Independent
States of the Former Soviet Union) under Contract 00-00587.

\begin{thebibliography}{99}
\bi{Ger} S. B. Gerasimov, Yad. Fiz. {\bf 2}, 598 (1965)
[Sov. J. Nucl Phys. {\bf 2}, 430(1966)].
\bi{DH} S. D. Drell and A. C. Hearn, Phys. Rev. Lett. {\bf 16}, 908 (1966).
\bi{E143} E143 Collaboration, K. Abe et al., Phys. Rev. Lett. {\bf78}, 815 (1997).
\bi{JLAB}E94010 Collaboration, M. Amarian {\it et al.}, Phys. Rev. Lett. {\bf 89}, 242301 (2002).
\bi{Hermes}HERMES Collaboration, A. Airapetian
{\it et al.}, 
Phys. Lett. {\bf B494}, 1 (2000);
Eur. Phys. J. {\bf C26}, 527 (2003).
\bi{JLAB2} CLAS Collaboration, J. Yun {\it et al.}, Phys. Rev. {\bf C67}, 055204 (2003) and
R. Fatemi {\it et al.}, Phys. Rev. Lett. {\bf 91}, 222002 (2003).
\bi{ST93} J. Soffer and O. Teryaev, Phys. Rev. Lett. {\bf 70}, 3373 (1993).
\bi{ST95} J. Soffer and O. Teryaev, Phys. Rev. {\bf D51}, 25 (1995).
\bi{KatSid} A. L.~Kataev, G. Parente and A. V. Sidorov,
Nucl.\ Phys.\ {\bf B573}, 405 (2000).
\bi{BC} H. Burkhardt and W. N. Cottingham, Ann. Phys. (N.Y.) {\bf 16}, 543 (1970).
\bi{Sch} J. Schwinger, Proc. Nat. Acad. Sci. U.S.A. {\bf 72}, 1559 (1975).
\bi{ST02}  J. Soffer and O. Teryaev, Phys. Lett. {\bf B545}, 323 (2002).
\bi{Ioffe} V. D. Burkert and B. L. Ioffe, Phys. Lett. {\bf B296}, 223 (1992) and
J.\ Exp.\ Theor.\ Phys.\  {\bf 78}, 619 (1994) [Zh. Eksp. Teor. Fiz. {\bf 105}, 1153 (1994)];
V.D. Burkert and Z. Li Phys. Rev. {\bf D47}, 46 (1993).
V. D. Burkert, AIP Conf. Proc. {\bf 603}, 3 (2001) and Nucl. Phys. {\bf A699}, 261 (2002).
\bi{Power} V.~M.~Braun and A.~V.~Kolesnichenko,
Nucl. Phys. {\bf B283}, 723 (1987).
\bi{simula}
S.~Simula, M.~Osipenko, G.~Ricco and M.~Taiuti,
Phys. Rev. {\bf D65}, 034017 (2002)
 \bi{Pert} S.~A.~Larin, F.~V.~Tkachov and J.~A.~Vermaseren,
Phys.\ Rev.\ Lett.\  {\bf 66}, 862 (1991);
S.~A.~Larin and J.~A.~Vermaseren,
Phys. Lett.  {\bf B259}, 345 (1991).
\bi{book} B. L. Ioffe, V. A. Khoze and L. N. Lipatov, {\it Hard Processes},
(North-Holland, Amsterdam, 1984).
Nucl. Phys. {\bf B283}, 723 (1987).
\bi{sol} N.~Y.~Lee, K.~Goeke and C.~Weiss,
Phys.\ Rev.\ {\bf D65}, 054008 (2002).
\bi{Mad} R. Madey {\it et al.}, Phys.\ Rev.\ Lett.\  {\bf 91}, 122002 (2003).
\bi{Deur} A. Deur {\it et al.}, arXiv:hep-ex/0407007.
\bi{ot03} O. V. Teryaev, arXiv:hep-ph/0303003.
\bi{aren} H. Arenh\"ovel, arXiv:nucl-th/0006083.
\bi{dodge} G. Dodge, talk at GDH2004, June 2-5 2004, Old Dominion University.
\bi{ji} X.~D.~Ji and W.~Melnitchouk, Phys.\ Rev.\  {\bf D56}, 1 (1997).

\eb

\end{document}